\documentclass[%
 reprint,
superscriptaddress,
 amsmath,amssymb,
prl
]{revtex4-2}

\usepackage{graphicx}
\usepackage{dcolumn}
\usepackage{bm}

\begin{document}


\title{Solitary Spectrally-Discrete Bound States in the Continuum in an Open System}

\author{Samyobrata Mukherjee}
\author{Jordi Gomis-Bresco}
\affiliation{ICFO-Institut de Ci\`{e}ncies Fot\`{o}niques, The Barcelona Institute of Science and Technology,  08860 Castelldefels (Barcelona),  Spain.
}

\author{David Artigas}
 \affiliation{ICFO-Institut de Ci\`{e}ncies Fot\`{o}niques, The Barcelona Institute of Science and Technology,  08860 Castelldefels (Barcelona),  Spain.
}
\affiliation{Department of Signal Theory and Communications, Universitat Polit\`{e}cnica de Catalunya,  08034 Barcelona, Spain
}%

\author{Lluis Torner}
\email{lluis.torner@icfo.eu}
\affiliation{ICFO-Institut de Ci\`{e}ncies Fot\`{o}niques, The Barcelona Institute of Science and Technology,  08860 Castelldefels (Barcelona),  Spain.
}
\affiliation{Department of Signal Theory and Communications, Universitat Polit\`{e}cnica de Catalunya,  08034 Barcelona, Spain
}%


\begin{abstract}
\noindent Bound states in the continuum (BICs) exist in a variety of physical systems where they appear as lossless propagating states surrounded by radiating modes. However, in the case of open systems, they coexist with continuous families of guided states, which may be modes or other BICs, located in different regions of the frequency-momentum parameter space. Here we report that waveguiding structures comprising anisotropic materials with two radiation channels where continuous families of modes are not possible whatsoever can support discrete BICs that exist for a single frequency and a single propagation direction. The results are isolated spectrally-discrete BIC states existing as lossless needles emerging from a sea of radiating leaky modes. 
\end{abstract}

\maketitle

\noindent Bound states in the continuum are modes that propagate without losses even though their existence domain overlaps with the part of the parameter space that corresponds to radiating waves. They were predicted in the context of quantum mechanics by von Neumann and Wigner \cite{Neuman1929} and later extended by Stillinger and Herrick \cite{Stillinger1975}. Subsequently, Friedrich and Wintgen described BICs as a general wave phenomenon \cite{Friedrich1985}, and since then they have been observed in many physical settings, including acoustic systems \cite{Parker1966}, quantum systems \cite{Kim1999} and photonic systems, where they have inspired a wealth of new phenomena as well as prospects of important applications \cite{Hsu2016, Koshelev2019_rev}. Photonic BICs were proposed in dielectric gratings \cite{Marinica2008} and in photonic crystal waveguides \cite{Bulgakov2008} and observed in waveguide arrays  \cite{Plotnik2011,Corrielli2013} and in photonic crystal slabs \cite{Hsu2013}. These studies spurred research into many systems, such as hybrid plasmonic-photonic systems \cite{Azzam2018}, zero-index materials \cite{Li2017, Minkov2018}, diffraction gratings \cite{Monticone2017}, anisotropic structures \cite{Gomis-Bresco2017}, all-dielectric \cite{Fan2019, Han2019} and plasmonic \cite{Liang2020} metasurfaces and several periodic systems \cite{Bulgakov2014, Bulgakov2015, Bulgakov2017a, Cerjan2019, Ovcharenko2020}, a photonic crystal with a liquid-crystalline anisotropic layer \cite {Pankin2020}, among others. In addition to being of fundamental interest, the existence of photonic BICs has been harnessed to realise applications in lasers \cite{Kodigala2017,Midya2018}, biosensing \cite{Romano2018}, multiplexed communication channels \cite{Yu_2020}, directional radiation \cite{Rivera2016, Yin2020}, broadband light capture \cite{Hayran2020}, high Q resonators \cite{Rybin2017,Jin2019}, or nonlinear optics \cite{Carletti2018, Liu2019}, to cite a few examples.

In the above-mentioned settings involving open waveguiding systems, BICs always coexist with bands of guided modes, albeit both types of states have a different origin and occupy different parts of the parameter space. For example, photonic crystal slabs, placed in a geometry that is symmetric along the axis perpendicular to the periodic layer, can support propagating interference BICs in the radiation bands originated by the destructive interference of the radiation channel, which coexist with both BICs in the $\Gamma$-point protected by symmetry and, importantly, with branches of standard guided Bloch modes that exist below the light cone \cite{Hsu2013}.  

In contrast, in this Letter we uncover that anisotropy-induced BICs arising in waveguiding structures with multiple radiation channels containing uniaxial materials may allow the existence of one, or several, discrete BICs that are {\it the sole bound state possible in the whole parameter space available to the system\/}. Therefore, such BICs appear as isolated, needle-like spectrally-discrete states in the frequency-momentum dispersion diagram, thus they resemble the discrete resonances that occur in closed systems.

\begin{figure}[t]
    \centering
    \includegraphics[width=1 \linewidth]{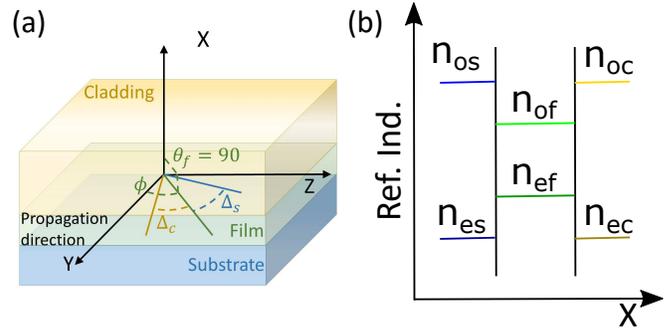}
    \caption{Figure 1 (a) Layout of the system. The yellow, green, and blue lines show the optical axis (OA) for the cover, film and substrate, respectively. The coordinate system is aligned along the wave propagation direction $y$, and $x$ is perpendicular to the interface. (b) Schematic of the refractive indices showing three uniaxial materials with negative birefringence.}
    \label{fig:1}
\end{figure}

The simplest structure supporting needle-like BICs is a planar {\it antiguiding\/} structure where all layers are made of negative birefringent media, with all optical axes (OAs) parallel to the interface plane ($\theta_c = \theta_f = \theta_s = 90^{\circ}$) and having identical  materials in the cladding and substrate [Fig. \ref{fig:1}(a)]. Antiguiding is ensured when the ordinary refractive index of the cladding, $n_{oc}$, and substrate, $n_{os}$ are the highest ones in the structure, so that in the cladding and substrate $n_{oc}=n_{os}>n_{ec}=n_{es}$, and in the film  $n_{os}>n_{of}>n_{ef}>n_{es}$, as shown in Fig. \ref{fig:1}(b). As the cladding and substrate ordinary refractive indices are the highest ones in the structure and are independent of the OA orientation, only semi-leaky modes are supported. Semi-leaky modes are improper modes that provide a good approximation of the field near the waveguide and are characterized by a complex mode index $N$, where $\emph{Im} \{N\}$ approximates the radiation loss \cite{Hu2009} and $n_{ec}=n_{es}< \emph{Re} \{N\}<n_{os}=n_{oc}$. Therefore, the radiation channels in the cover and substrate are the ordinary wave, which couples to the radiation continuum, while the extraordinary wave is evanescent \cite{Marcuse1979, Knoesen1988, Torner1993}. The coordinate axes are centered at the substrate/film interface, where $x$ is normal to the interface plane, $y$ is the propagation direction and $\phi$ gives the angle between the film OA and the direction of propagation. To have independent control over the cladding and substrate radiation channels, their OAs are free to rotate independently in the interface plane. $\Delta_{c/s} = \phi_{c/s}-\phi$ gives the offset between the film OA and the cover/substrate OAs. Then, the extraordinary and ordinary electric field in the cladding and substrate can be written as
\begin{equation}
    \begin{split}
        E_{c,p}^e+E_{c,p}^o =& \left(A_{c,p}^e e^{k_0 \gamma_{e}^c (x+D)} + A_{c,p}^o e^{k_0\gamma_{o}^s (x+D)}\right) e^{- i N k_0 y}\\
        E_{s,p}^e+E_{s,p}^o =& \left(A_{s,p}^e e^{-k_0 \gamma_{e}^s x} + A_{s,p}^o e^{- k_0\gamma_{o}^s x}\right) e^{- i N k_0 y}.
     \end{split}
     \label{fields}
\end{equation}
\noindent Here, $p=x,y$ or $z$ represents the field components, $k_0$ is the free space wavenumber, $D$ is the thickness of the film and the transverse ordinary ($\gamma_o^{i}$) and extraordinary ($\gamma_e^{i}$) decay constants are given by
\begin{equation}
\begin{split}
    \gamma_o^{i}  = & \pm \sqrt{N^2-n_{oi}^2}, \\
    \gamma_e^{i}  = & \pm \sqrt{N^2 \left((n_{ei}^2/n_{oi}^2) \cos^2{\phi_{i}}+\sin^2{\phi_{i}} \right)-n_{ei}^2},
\end{split}
    \label{gammas}
\end{equation}
\noindent where $N$ is the mode index, $n_{e/o}$ stands for the extraordinary and ordinary refractive indices and $i=c$ or $s$ is either the cladding or the substrate. The extraordinary and ordinary field in the film can be written as
\begin{equation}
\begin{split}
    E_{f,p}^e & + E_{f,p}^o = [ B_{f,p}^e \sin{\left (k_0\kappa_{e} x \right )} + B_{f,p}^o \sin{\left (k_0\kappa_{o} x \right )} \\
     & + C_{f,p}^e \cos{\left (k_0\kappa_{e} x \right )} + C_{f,p}^o \cos{\left (k_0\kappa_{o} x \right )} ] e^{- i N k_0 y},
\end{split}
     \label{fieldf}
\end{equation}
where the transverse wavevectors $\kappa_{e/o}$ for the ordinary and extraordinary polarization are given by
\begin{equation}
\begin{split}
    \kappa_o  = & \sqrt{n_{of}^2 - N^2}, \\
    \kappa_e  = &\sqrt{n_{ef}^2 - N^2 \left((n_{ef}^2/n_{of}^2) \cos^2{\phi}+\sin^2{\phi} \right)}.
\end{split}
\end{equation}

\begin{figure}[t]
    \includegraphics[width=1 \linewidth]{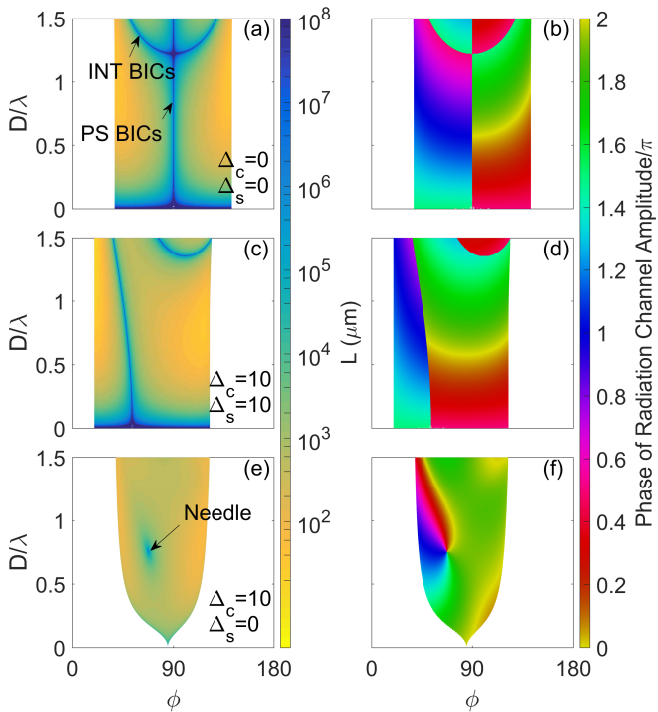}
    \caption{BICs embedded on semi-leaky modes and corresponding phase maps. (a, c, e) Propagation length $L$, defined as the length at which the field amplitude decays to $1/e$ of the initial value for the fundamental semi-leaky mode. (b, d, f)  corresponding phase of the cladding radiation channel (ordinary) amplitude, measured with respect to the confined (extraordinary) wave. The OA orientations are (a,b) $\Delta_c=\Delta_s=0^{\circ}$ (c,d) $\Delta_c=\Delta_s=10^{\circ}$ and (e,f) $\Delta_c=10^{\circ}, \Delta_s=0^{\circ}$. The phase in the substrate radiation channel is obtained by adding $\pi$ to the phase of the cladding radiation channel. The transition from the colored sheet to white corresponds to the leaky mode cut-off.}
    \label{fig:2}
\end{figure}

Applying the boundary conditions at $x=0$ and $x=D$ gives us the dispersion equation for the structure which can be written in a compact form as
\begin{equation} 
    W \left(N, D/\lambda, \nu, \Phi \right)=0,
    \label{dispersion}
\end{equation}
where $D/\lambda$ is the normalized thickness or frequency, $\nu$ and $\Phi$ represent the set of all refractive indices and OA angles of the system, respectively. Selecting the negative square root in eq. (\ref{gammas}) for the two radiation channels, $\gamma_o^{c/s}$  gives the complex, semi-leaky mode solutions of the dispersion equation (\ref{dispersion}). Using the Berreman transfer matrix formulation \cite{Berreman1972, Gomis-Bresco2017, Mukherjee2018}, the fields in the substrate and the cladding can be related as
\begin{equation}
	a^o_s\cdot\vec{v}^o_s + a^e_s\cdot\vec{v}^e_s = a_c^{o}\cdot\hat{M}\vec{v}_c^{o} + a_c^{e}\cdot\hat{M}\vec{v}_c^{e},
	\label{amplitudes}
\end{equation}

\noindent where $v^j_i=[E_{i,y}^j, z_0 H_{i,z}^j, E_{i,z}^j, z_0 H_{i,y}^j, ]^T $ is a column vector containing the tangential field amplitudes of the basis wave in the medium, $a^j_i$ is the complex amplitude of the corresponding basis wave, $z_0$ is the vacuum impedance,  and $\hat{M}$ is the transfer matrix for the film. For a BIC to exist, the coefficients of the radiation channels corresponding to the ordinary polarisation have to vanish, i.e., $a^o_c=a^o_s=0$ in eq. (\ref{amplitudes}), which reduces to 
\begin{equation}
	a^e_s\cdot\vec{v}^e_s - a_c^{e}\cdot\hat{M}\vec{v}_c^{e}=0.
	\label{amplitudes_reduced}
\end{equation}
The condition for BIC existence can thus be set by requiring that solutions simultaneously exist for Eqs.~(\ref{dispersion}) and (\ref{amplitudes_reduced}). Without loss of generality, we set the cladding and substrate refractive indices $n_{ec}=n_{es}=1.3$ and $n_{oc}=n_{os}=1.7$, and film refractive indices $n_{ef}=1.4$ and $n_{of}=1.6$. When the structure has full anisotropy-symmetry \cite{Mukherjee2018}, i.e., the optic axes of the cover, film and substrate are aligned ($\Delta_c=\Delta_s=0^{\circ}$) and in the interface plane ($\theta_c=\theta_f=\theta_s=0^{\circ}$), the cladding and substrate radiation channels are equivalent and the semi-leaky mode supports families of both polarisation separable (PS) and interference (INT) BICs, which appear as lossless lines within the leaky mode sheet in the $\phi-D/\lambda$ space. This is shown in Fig. \ref{fig:2}(a), which shows the $1/e$ propagation distance $L$ for the fundamental semi-leaky mode. PS BICs are analogous to symmetry protected BICs in other photonic structures \cite{Hsu2016} and exist at $\phi=90^{\circ}$, corresponding to the guiding profile created by the extraordinary indices shown in Fig. \ref{fig:1}(b). INT BICs are hybrid, full-vector modes that arise from destructive interference simultaneously in both radiation channels. As the anisotropy symmetry is maintained and the cladding and substrate radiation channels are equivalent, the fundamental semi-leaky mode supports a curved line of existence of INT BICs [Fig. \ref{fig:2}(a)]. Since BICs are zeroes of radiation, the phase of the radiation channel amplitude at a BIC is undefined and BIC existence loci correspond jumps of $\pm\pi$ in the phase of the radiation channel amplitude \cite{Mukherjee2018}. This is shown in Fig. \ref{fig:2}(b), where the phase in the cladding radiation channel is measured with respect to the phase of the confined wave, i.e., the extraordinary polarisation. The jumps in the phase of the substrate radiation channel at the BIC lines loci is opposite in sign (adding $\pi$) to the jump shown in Fig. \ref{fig:2}(b).

Breaking the azimuthal anisotropy-symmetry, but maintaining the geometric symmetry of the structure keeping identical OA orientations in the cladding and substrate ($\Delta_c=\Delta_s \neq 0^{\circ}$), maintains the equivalence between the cover and substrate radiation channels, and the structure still supports lines of existence of BICs. However, all the lines are deformed and correspond to INT BICs only, as shown in Fig. \ref{fig:2}(c,d) for $\Delta_c=\Delta_s=10^{\circ}$. Thus, this is weak anisotropy-symmetry breaking \cite{Mukherjee2018}. 

\begin{figure}[t]
    \centering
    \includegraphics[width=1 \linewidth]{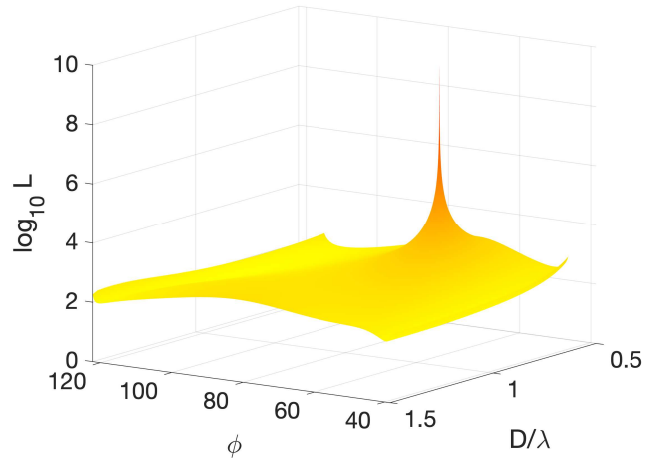}
    \caption{Propagation distance $L$ as a function of $\phi$ and $D/\lambda$ in 3D. The needle-like BIC is a lossless point surrounded by the radiating semi-leaky mode within which it is embedded.}
    \label{fig:3}
\end{figure}

Different OA orientations in the cladding and the substrate mean that, in addition to the anisotropy symmetry, the geometrical symmetry in the structure is also broken and the radiation channels are no longer equivalent. This situation is shown in Fig. \ref{fig:2}(e) for the fundamental semi-leaky mode, where the OA in the substrate is aligned with the OA in the film ($\Delta_s=0^{\circ}$) but the cladding OA has an offset $\Delta_c=10^{\circ}$. As a result, the we see a topological transition in the map of the mode propagation distance $L$ on the fundamental semi-leaky mode, and the BIC lines of existence shown in \ref{fig:2}(a,b) collapse to an isolated BIC point [Fig. \ref{fig:2}(e)]. The isolated BIC corresponds to a zero in both radiation channels and results in a phase singularity in the radiation channel amplitude characterised by a screw phase dislocation, as shown in Fig. \ref{fig:2}(f). The winding number assigned to the BIC according to the sense of the screw phase dislocation in the phase of the radiation channel has opposite sign in the cladding and substrate, $-1$ (anti-clockwise increase) and $+1$ (clockwise increase), respectively. This  BIC point, isolated in both wavelength and direction, is the only bound mode supported by the structure with these OA orientations (for these values of $\phi, \Delta_c$ and $\Delta_s$) and in a broad range of frequencies $D/\lambda$. Since anisotropy-symmetry is broken, this structure cannot support PS BICs and the BIC in Fig. \ref{fig:2}(e,f) is an INT BIC with hybrid polarization. Fig. \ref{fig:3} shows how the propagation length $L$ diverges, demonstrating the resemblance of the BIC to an isolated, lossless needle in an environment of radiating semi-leaky modes.

\begin{figure}[t]
    \centering
    \includegraphics[width=1 \linewidth]{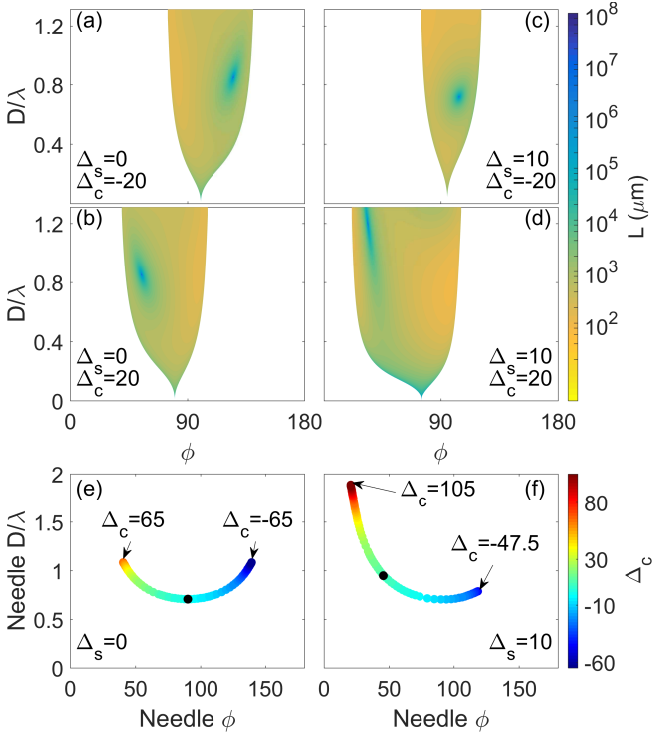}
    \caption{Leaky mode sheet for  $\Delta_s=0^{\circ}$ and (a) $\Delta_c=- 20^{\circ}$ (b) $\Delta_c=+ 20^{\circ}$. The leaky mode sheet in (c) and (d) has the same values of $\Delta_c$ but with $\Delta_s=10^{\circ}$. Locus of the needle in the $\phi-D/\lambda$ under variation of $\Delta_c$ for fixed values of (e) $\Delta_s=0^{\circ}$ and (f) $\Delta_s=10^{\circ}$. The color  of the dots in (e) and (f) indicate the value of $\Delta_c$. The black dot on each plot corresponds to the scenario $\Delta_c=\Delta_s$ where the radiation channels are equivalent and instead of the needle the structure supports BIC lines of existence, as shown in Figs. \ref{fig:2}(a) and (c).}
    \label{fig:4}
\end{figure}

Since the isolated BIC (needle) can be assigned winding numbers, it is robust under variation of OA orientation, which only changes its position on the semi-leaky mode. This is shown in Fig. \ref{fig:4} which presents the position of the needle within the fundamental semi-leaky mode sheet under variation of OA orientation. The first point to note when comparing Figs. \ref{fig:4}(a) to (b) is that setting $\Delta_c \neq 0^{\circ}$ breaks the symmetry of the leaky mode sheet around $\phi=90^{\circ}$ and determines the position of the needle. In addition, $\Delta_s$ has a substantial impact on the leaky mode cutoff, as can be seen by comparing Fig. \ref{fig:4}(a) with (c) and Fig. \ref{fig:4}(b) with (d). In general we see that the greater the difference between $\Delta_c$ and $\Delta_s$ the narrower the $\phi$-range of existence of leaky modes. It is apparent from comparing the four Figs. \ref{fig:4}(a-d) that the needle can be positioned in a broad range of frequencies $D/\lambda$ and propagation directions $\phi$ by tuning the substrate and cladding OA orientations. Fig. \ref{fig:4}(e) shows the locus of the needle in the $\phi-D/\lambda$ space as a function of variation of $\Delta_c$ with $\Delta_s=0^{\circ}$ while Fig. \ref{fig:4}(f) shows the locus of the needle for $\Delta_s=10^{\circ}$. Fig. \ref{fig:4}(e) shows that when $\Delta_s=0^{\circ}$ the locus of the needle is symmetric about $\phi=90^{\circ}$. The needle exists for a range of values $-65^{\circ}<\Delta_c<65^{\circ}$. When $\Delta_s=10^{\circ}$, the symmetry of the locus is broken and the needle exists for a greater range of values of $-47.5^{\circ}<\Delta_c<105^{\circ}$. The range of existence of the needle in $D/\lambda$ is also larger, while the range of existence in $\phi$ is shifted to lower values but is similar in extent to the case when $\Delta_s=0^{\circ}$. The needle only stops existing when it moves beyond the cutoff of the fundamental semi leaky mode on which it exists (extreme values of $\Delta_c$), showing its robustness against perturbations.

There are other needles that occur at different values of $\phi$ and $D/\lambda$, both on the fundamental semi-leaky mode and on higher order leaky modes. The needle shown in Fig. \ref{fig:2}(e,f) occurs at $D/\lambda=0.758$ while the next needle on the fundamental (zero order) leaky mode occurs at $D/\lambda=2.162$ (not shown), almost two octaves beyond. In the case of the first order leaky mode, the needle appears even beyond, for $D/\lambda=2.682$. Therefore, the needle shown in Fig. \ref{fig:2}(e) is the only bound mode supported by the structure for any practical operation range of wavelengths. 

The transition of BIC lines of existence to BIC points in our structure is related to the strong anisotropy-symmetry breaking mechanism introduced in  \cite{Mukherjee2018}. There, the polar anisotropy-symmetry was broken by taking the film OA out of the interface plane ($\theta_f \neq 90^{\circ}$) in a structure with only one radiation channel. As a result, the mismatch between the polarization of the wave in the film and the radiation channel prevented the effective destructive interference, resulting in as many BIC points as BIC lines were present in the fully anisotropy-symmetric structure, with very close values of $D/\lambda$. In the phenomenon discovered here the transition from BIC lines to points has a fundamentally different nature, since polar anisotropy-symmetry is not broken. The mechanism preventing BIC lines of existence is the different OA orientations in the cladding and the substrate, which result in two distinct radiation channels. This leads to an additional constraint being placed on BIC existence and a consequent drop in dimension of the BIC solution from lines to points \cite{Hsu2016}. However, this is more complex than the simple interpretation of finding the point of intersection of independent solutions corresponding to each radiation channel, since the radiation channels are linked by reflections as encapsulated in the transfer matrix in eq. (\ref{amplitudes}). Physically, the amount of radiation in each channel is modified due to simultaneous breaking of the geometrical and anisotropic symmetry, which destroys all the BIC lines except for one point.

In conclusion, we have found that spectrally isolated bound modes exist in anisotropic anti-guiding structures that do not support any other bound modes whatsoever. Such unique, needle-like modes exist under simultaneous azimuthal anisotropy-symmetry breaking and geometrical breaking of the symmetry, resulting in two distinct radiation channels. Here we have shown the effect in the simplest geometry, but we have verified that it also occurs in configurations with two channels and breaking of the polar anisotropy-symmetry. The isolated BIC needle is robust and its locus on the semi-leaky mode sheet can be tuned under variation of the OA orientation. We anticipate that analogous phenomena may occur in other open photonic systems featuring simultaneous directive and frequency filtering, which is an important property in lasing \cite{Kodigala2017, Midya2018}, while its sensitivity suggests applications in sensing \cite{Romano2018}.

\noindent {\bf Funding}. H2020 Marie Skłodowska-Curie Action GA665884; Government of Spain (grants PGC2018-097035-B-I00; Severo Ochoa CEX2019-000910-S);  Fundació Cellex; Fundació Mir-Puig; Generalitat de Catalunya (CERCA and AGAUR 2017-SGR-1400).

\bibliography{bibliography}

\end{document}